\documentclass[pdflatex,sn-nature]{sn-jnl}%

\usepackage{graphicx}%
\usepackage{multirow}%
\usepackage{amsmath,amssymb,amsfonts}%
\usepackage{amsthm}%
\usepackage{mathrsfs}%
\usepackage[title]{appendix}%
\usepackage{xcolor}%
\usepackage{textcomp}%
\usepackage{manyfoot}%
\usepackage{booktabs}%
\usepackage{algorithm}%
\usepackage{algorithmicx}%
\usepackage{algpseudocode}%
\usepackage{listings}%
\usepackage{soul} 
\usepackage{xr} 

\usepackage{multirow}
\usepackage{makecell}
\usepackage{pdflscape}
\usepackage[table]{xcolor}
\usepackage{rotating}
\usepackage{threeparttable}
\usepackage{booktabs}
\usepackage{tabularx}

\usepackage{geometry}
\theoremstyle{thmstyleone}%
\theoremstyle{thmstyletwo}%

\theoremstyle{thmstylethree}%

\begin{document}

\title[Article Title]{Decadal sink-source shifts of forest aboveground carbon since 1988}

\author*[1,2]{\fnm{Zhen} \sur{Qian}}\email{zhen.qian@tum.de}

\author[1,2]{\fnm{Sebastian} \sur{Bathiany}}

\author[1,2,3]{\fnm{Teng} \sur{Liu}}

\author[1,2]{\fnm{Lana L.} \sur{Blaschke}}

\author[1,2]{\fnm{Hoong Chen} \sur{Teo}}

\author*[1,2]{\fnm{Niklas} \sur{Boers}}\email{n.boers@tum.de}

\affil[1]{\orgdiv{Munich Climate Center and Earth System Modelling Group}, \orgname{Department of Aerospace and Geodesy, TUM School of Engineering and Design, Technical University of Munich}, \city{Munich}, \postcode{80333}, \country{Germany}}

\affil[2]{\orgname{Potsdam Institute for Climate Impact Research}, \city{Potsdam}, \postcode{14473}, \country{Germany}}

\affil[3]{\orgdiv{School of Systems Science and Institute of Nonequilibrium Systems}, \orgname{Beijing Normal University}, \city{Beijing}, \postcode{100875}, \country{China}}


\abstract{
Forest ecosystems are vital to the global carbon cycle, yet their long-term aboveground carbon (AGC) dynamics remain uncertain.
Here, we integrate multi-source satellite observations with probabilistic deep learning models to reconstruct a harmonized, uncertainty-aware global forest AGC record from 1988 to 2021 at 0.25$^\circ$.
We find that, although global forests sequestered 6.2 PgC, moist tropical and boreal forests have progressively transitioned toward carbon sources since the early 2000s.
This shift coincides with a strengthening negative correlation between tropical AGC variability and atmospheric CO$_\textnormal{2}$ growth rates ($\textnormal{r} = \textnormal{-0.63}$ in 2011--2021), suggesting tropical forests increasingly modulate the global carbon cycle.
Notably, in the Brazilian Amazon, the contribution of intact forests to the year-to-year variations in AGC losses increased from 33\% in the 1990s to 76\% in the 2010s, surpassing that of deforested areas (from 60\% to 13\%).
Our findings highlight the vulnerability of carbon stocks in key biomes and provide a benchmark to track emerging sink-source shifts under anthropogenic climate change.
}

\maketitle

\section*{Introduction}\label{sec1}

Forest ecosystems play a critical role in the terrestrial carbon cycle, storing approximately 45\% of terrestrial carbon and acting as long-term carbon sinks~\cite{bonanForestsClimateChange2008,panEnduringWorldForest2024}.
Aboveground carbon (AGC), which accounts for 30--50\% of total forest carbon pools~\cite{maGlobalDistributionEnvironmental2021,moIntegratedGlobalAssessment2023}, is a fundamental indicator of forest productivity and an essential climate variable for monitoring biospheric changes~\cite{plummerESAClimateChange2017}.
Under evolving climate change and disturbance regimes (e.g., fires, logging, and deforestation), forests can shift from being long-term sinks to sources of atmospheric $\text{CO}_2$, or vice versa~\cite{pughImportantRoleForest2019,wangDisturbanceSuppressesAboveground2021}.
Understanding such long-term, persistent shifts, beyond short-term natural variability driven by transient climate anomalies, helps reveal AGC dynamics associated with climate and land-use forcing and with long-term disturbance regimes.

Recent international initiatives, such as ESA’s Climate Change Initiative (CCI) and NASA’s Global Ecosystem Dynamics Investigation (GEDI), have improved the ability to map global and regional AGC stocks for recent years through combined field inventories and satellite remote sensing ~\cite{santoroDesignPerformanceClimate2024,dubayahGEDIL4BGridded2023}. 
However, these existing AGC maps typically provide only temporal snapshots or cover relatively short periods.
This temporal sparsity makes it challenging to comprehensively assess both long-term structural shifts and the interannual variability of AGC fluxes~\cite{fanSatelliteobservedPantropicalCarbon2019}. 

Therefore, microwave-based vegetation optical depth (VOD) measurements have emerged as a promising proxy for tracking vegetation biomass dynamics.
VOD, particularly at longer wavelengths such as the L-band, exhibits a near-linear spatial relationship with vegetation biomass~\cite{wigneronGlobalCarbonBalance2024}, and has substantially improved recent assessments of global and regional AGC variability~\cite{liuRecentReversalLoss2015,brandtSatellitePassiveMicrowaves2018,fanSatelliteobservedPantropicalCarbon2019,qinCarbonLossForest2021,fengDoublingAnnualForest2022,yangGlobalIncreaseBiomass2023,bar-onRecentGainsGlobal2025}.
However, L-band VOD records are only available from approximately 2010 onwards, restricting multi-decadal analyses and potential attribution of changes to anthropogenic climate change. 
In addition, emerging evidence suggests that VOD-derived AGC estimates can be influenced by soil moisture, which can matter particularly in low-biomass regions and can potentially lead to an overestimation of biomass fluctuations during transient water stress~\cite{koningsInterannualVariationsVegetation2021}.
Consistent with these limitations, there are substantial inconsistencies between existing AGC products (see Methods), even among products derived from similar data sources or methodological frameworks.
These discrepancies make it difficult to reconstruct long-term AGC dynamics by simply concatenating disjointed short-term records, and may compromise the internal consistency of a multi-decadal time series.
Therefore, there is a critical need for a robust, internally consistent, and multi-decadal global observational record to characterize long-term AGC dynamics across regions and through time.

Here, we address these critical knowledge gaps by integrating multi-source satellite and ground-based observations through a probabilistic deep learning approach.
We reconstruct spatio-temporally explicit AGC stocks and fluxes at 0.25\textdegree{} resolution from 1988 to 2021, together with prediction intervals that quantify uncertainties arising from both observational inputs and the modeling process.
Based on this probabilistic reconstruction, we conduct a systematic assessment of long-term AGC dynamics, examining spatiotemporal patterns and decadal-scale shifts between carbon sources and sinks across multiple spatial scales.
With a focus on tropical forests, particularly the Brazilian Amazon, we further examine regional AGC dynamics and their atmospheric consequences, highlighting the underlying interplay of natural and anthropogenic influences.
Collectively, our results provide a benchmark for tracking sink--source dynamics and interpreting long-term changes in global forest AGC under environmental and climate change.

\section*{Results}\label{sec2}

\subsection*{Reconstruction of global forest AGC}\label{subsec21}

We categorize remotely sensed vegetation variables and environmental data into dynamic (time-varying) and static (time-averaged or time-independent) predictors of AGC (see Methods; Supplementary Fig.1 and Supplementary Table 1).
Dynamic predictors include growing-season statistics of CXKu-band VOD, normalized difference vegetation index (NDVI), and leaf area index (LAI), as well as plant functional types (PFTs) of trees and forest cover fractions derived from land cover data.
Static predictors comprise aggregated L-band VOD, photosynthetically active radiation (PAR), land surface elevation, and geographic coordinates, which help represent broad spatial and biome-specific heterogeneity~\cite{langHighresolutionCanopyHeight2023}.

Using a probabilistic deep learning framework based on convolutional neural networks (CNNs)~\cite{langCountrywideHighresolutionVegetation2019,langHighresolutionCanopyHeight2023}, we model the spatial relationships between these predictors and the ESA CCI AGC reference maps~\cite{santoroDesignPerformanceClimate2024} (see Methods; Supplementary Fig. 1). 
To improve training robustness, our CNN explicitly incorporates per-grid-cell uncertainty from the ESA CCI AGC products into its loss function, enabling the model to weight observations based on their confidence.
Furthermore, to reduce the risk of overfitting to spatial patterns and learning spurious year-to-year fluctuations inherent in individual ESA CCI AGC snapshots (Supplementary Fig. 2), we train independent CNNs for each available reference year (2015--2020) and combine them into an ensemble.
This ensemble strategy, together with uncertainty quantification techniques~\cite{kendallWhatUncertaintiesWe2017,lakshminarayananSimpleScalablePredictive2017}, yields AGC intervals rather than only deterministic estimates.
These intervals represent uncertainties arising from both the inherent noise in the satellite data (aleatoric uncertainty) and the limitations of the model itself (epistemic uncertainty).
Finally, we apply the trained ensemble to reconstruct a continuous, harmonized time series of global AGC maps from 1988 to 2021.

We investigate the contributions of predictors to the CNNs' predictions using an Explainable AI (XAI) approach. 
Specifically, we aggregate feature attributions based on integrated gradients~\cite{sundararajan2017axiomatic} across CNN ensemble members to quantify predictor importance (Supplementary Fig. 3). 
This analysis reveals that dynamic predictors account for approximately 56\% of total attribution, indicating that time-varying vegetation signals play a major role in the reconstruction.
Static biophysical and environmental variables (including PAR, elevation, and L-band VOD) contribute about 26\%, leaving static encoded geographic coordinates ($<19$\%) to serve as a complementary spatial context for these dominant eco-physiological drivers.

To evaluate its predictive performance, we benchmark our model against conventional empirical VOD-to-AGC conversions and classical machine learning methods (Table~\ref{tab:model_comparison}).
While VOD is a valuable biomass proxy, empirical models relying on VOD alone exhibit limited predictive capability on held-out ESA CCI test data from 2010 and 2021 ($R^2 \approx 0.2\text{--}0.6$), and their estimated time series of global annual total AGC show weak correlation with the independent 2000--2019 reference records~\cite{xuChangesGlobalTerrestrial2021} ($r=0.35,\,p=0.13$). 
In contrast, integrating multi-source data via our probabilistic CNNs improves both predictive performance ($R^2=0.97$) and agreement of the global total AGC time series ($r=0.70,\,p<0.001$).
Compared with classical machine-learning models using the same multi-source predictors, linear models (Lasso and ridge) perform worse in both space and time, whereas random forest achieves comparable predictive performance ($R^2=0.98$) but lower temporal consistency ($r=0.45,\,p<0.05$).

We further compare the reconstructed AGC maps with multiple independent remotely sensed AGC references (Supplementary Table 2 and Supplementary Note 1).
Globally, our reconstructions demonstrate strong spatiotemporal agreement with the AGC references, with grid-cell-wise spatial correlations up to 0.87 and temporal correlations of regionally aggregated AGC stocks up to 0.70 (Table~\ref{tab:global_agc_r} and Supplementary Fig.4).
Although this spatial agreement decreases when evaluated within individual biomes, and temporal agreement drops for interannual AGC fluxes compared to total stocks, cross-comparisons reveal even more pronounced discrepancies among the reference datasets themselves due to differing data sources and methodologies (Supplementary Figs. 6, 7 and 8).
Notably, compared to these reference products, our AGC estimates tend to correlate more strongly with each individual dataset than these datasets correlate with each other.
This indicates that our new observational AGC record provides a useful, internally consistent benchmark for long-term AGC assessments.

Beyond comparisons with satellite-derived AGC references, we compare our reconstruction with national inventory data of living biomass carbon (i.e., AGC plus belowground carbon) from Pan et al.~\cite{panEnduringWorldForest2024}.
Following their biome definitions and temporal intervals (used here only for benchmarking), we find high agreement in regional carbon stocks (Fig.~\ref{fig:fig1}a).
When evaluating net changes in carbon stocks at decadal scales (Fig.~\ref{fig:fig1}b), our estimates generally fall within the spread of values reported by other independent satellite-derived products for the 2000s and 2010s.
Furthermore, across boreal and temperate regions, the temporal trajectories of our estimated carbon changes generally align with the inventory-based estimates from Pan et al.~\cite{panEnduringWorldForest2024}.
We note that the estimated magnitudes from our model, as well as those from other satellite-derived products, tend to be lower than the inventory data.
This discrepancy likely arises from differences in the target variables: assessments by Pan et al.~\cite{panEnduringWorldForest2024} consider total living biomass carbon, while ours focuses on AGC.
Larger discrepancies between satellite-based estimates and inventory data occur primarily in tropical regions.
One likely reason is that ground-based inventory coverage is sparse in the tropics, so inventory assessments rely more heavily on bookkeeping approaches to represent losses associated with deforestation and degradation~\cite{yangGlobalIncreaseBiomass2023}.
In this context, the contrast between satellite observations and inventories underscores the value of our continuous long-term AGC record as a complement to sparse ground monitoring networks, particularly during the observationally limited 1990s.

\subsection*{Spatiotemporal dynamics of AGC stocks and fluxes}\label{subsec22}

We use our reconstruction to characterize the global distribution and long-term dynamics of forest AGC.
Spatially, our estimates reveal high AGC densities in equatorial forests that decline progressively toward higher latitudes (Fig.~\ref{fig:fig2}a,c), consistent with independent references~\cite{liuRecentReversalLoss2015,avitabileIntegratedPantropicalBiomass2016,santoroDesignPerformanceClimate2024,dubayahGEDIL4BGridded2023,fanSatelliteobservedPantropicalCarbon2019,boitardAbovegroundBiomassDataset2025}.
The uncertainty (i.e., the standard deviation representing both data uncertainty and the model uncertainties captured by year-to-year variability; see Methods) is similarly elevated in dense tropical forests (Fig.~\ref{fig:fig2}b), possibly reflecting higher observational noise and lower predictive skill of CNNs in these areas.
Conversely, the relative uncertainty, defined as the ratio of the estimated uncertainty to the estimated AGC, is higher in low-biomass regions; even small deviations in estimated AGC in these areas can lead to disproportionately high relative uncertainty ratios (Supplementary Fig. 9).
Categorizing these long-term averages reveals that moist tropical forests dominate global stocks (119 PgC, 52\%; median $113\,\text{MgC\,ha}^{-1}$), followed by temperate (52 PgC, 23\%; median $37\,\text{MgC\,ha}^{-1}$), dry tropical and subtropical (35 PgC, 15\%; median $18\,\text{MgC\,ha}^{-1}$), and boreal forests (24 PgC, 11\%; median $17\,\text{MgC\,ha}^{-1}$) (Fig.~\ref{fig:fig2}d,e).

Over the period 1988--2021, grid-cell-wise AGC densities exhibit an overall increasing trend (median $0.07\,\text{MgC\,ha}^{-1}\,\text{yr}^{-1}$), driven primarily by temperate and boreal forests, while moist tropical forests experienced an overall carbon loss with a median trend of $-0.04\,\text{MgC\,ha}^{-1}\,\text{yr}^{-1}$ (Fig.~\ref{fig:fig3}a, e and Fig.~\ref{fig:fig4}a). 
In total, our AGC reconstruction shows that global forests sequestered a net 6.20 PgC in 1988--2021 (Fig.~\ref{fig:fig3}f). 
However, this overall carbon gain masks regional differences (Fig.~\ref{fig:fig4}e; Supplementary Table 3).
Specifically, temperate forests contributed 3.10 PgC (the largest share of the overall gain), followed by 1.96 PgC from dry tropical \& subtropical, and 1.25 PgC from boreal forests, whereas moist tropical forests acted as a weak source of -0.11 PgC.
The widespread AGC sequestration is likely driven by elevated atmospheric $\mathrm{CO}_2$ and, to some extent, nitrogen deposition~\cite{terrerNitrogenPhosphorusConstrain2019, walkerIntegratingEvidenceTerrestrial2021}. 
Conversely, deforestation, degradation, and climate change may counteract such benefits in the moist tropics~\cite{coxSensitivityTropicalCarbon2013,bacciniTropicalForestsAre2017,qinCarbonLossForest2021,lapolaDriversImpactsAmazon2023}.

We observe a substantial decrease in AGC during 1991 $(-3.16\,\text{PgC})$, with 92\% of this decrease occurring in tropical and subtropical biomes $(-2.92\,\text{PgC})$.
Two major factors may have contributed to this decrease: (\emph{i}) the compound climate stresses caused by the El~Ni\~no/Southern Oscillation (ENSO) and the Mount Pinatubo eruption, as well as the reduction in photosynthetically available radiation after the eruption, which could have caused widespread vegetation mortality~\cite{luchtClimaticControlHighLatitude2002,guPrecipitationTemperatureVariations2011,wangVariationsAtmosphericCO22013}, and (\emph{ii}) volcanic aerosols interfering with remote-sensing data, potentially introducing systematic biases into AGC retrievals~\cite{vargasEmpiricalNormalizationEffect2009}.

Investigating our AGC fluxes for decadal intervals (i.e., 1988--2000, 2001--2010, and 2011--2021) reveals pronounced temporal variability, probably driven by episodic climate extremes and shifting anthropogenic pressures (Fig.~\ref{fig:fig3}b-d, Fig.~\ref{fig:fig4}b-d, and Supplementary Table 3).
We quantify these dynamics using two complementary metrics: (i) the net change (i.e., sink or source) computed as the difference between the last and the first year of a decade~\cite{bacciniTropicalForestsAre2017,xuChangesGlobalTerrestrial2021}, and (ii) linear stock trends derived from the Theil–Sen estimator (Methods).
The reliability of these metrics is supported by Signal-to-Noise Ratio (SNR) evaluations and Monte Carlo-derived uncertainty ranges (Supplementary Figs. 10 and 11, Supplementary Tables 3 and 4).
Although grid-cell-level flux estimates are inherently susceptible to high-frequency natural variability, our SNR analysis shows that spatial aggregation at regional and global scales averages out such local noise, yielding robust decadal assessments.

Overall, global forests have remained a carbon sink: Decadal changes are $210.1\,\text{TgC\,yr}^{-1}$ during 1988--2000, $43.0\,\text{TgC\,yr}^{-1}$ during 2001--2010, and $196.3\,\text{TgC\,yr}^{-1}$ during 2011--2021, with positive values indicating carbon uptake by forests. 
Concurrently, decadal stock trends have continuously increased, with estimated rates of 105.5, 125.5, 265.9 $\text{TgC\,yr}^{-1}$ for the three respective periods, further confirming the enduring and even strengthening AGC sink of global forests.
However, these global trends mask pronounced decadal sink-to-source shifts in moist tropical and boreal forests (Fig.~\ref{fig:fig4} and Supplementary Table 3).

Moist tropical forests acted as a substantial carbon sink ($90.2\,\text{TgC\,yr}^{-1}$) during 1988--2000, but turned into a massive carbon source ($-190.6\,\text{TgC\,yr}^{-1}$) during 2001--2010.
This transition coincided with a period marked by repeated extreme events, such as droughts~\cite{phillipsDroughtSensitivityAmazon2009,lewis2010AmazonDrought2011} and fires~\cite{harrisonGlobalImpactIndonesian2009}, as well as accelerated deforestation prior to 2004~\cite{kimAcceleratedDeforestationHumid2015}, which has likely contributed substantially to the observed carbon losses.
Similar sink-to-source shifts were also observed during 2001--2010 in countries with extensive moist tropical forests, including Brazil, Indonesia, and Peru (Supplementary Table 3).
In the following decade (2011--2021), moist tropical forests transitioned toward a weak carbon source according to our dataset ($-4.7\,\text{TgC\,yr}^{-1}$).
This partial recovery is likely due to reductions in deforestation rates~\cite{arimaPublicPoliciesCan2014} and regrowth in previously cleared or degraded areas~\cite{poorterMultidimensionalTropicalForest2021}, although the specific causal drivers of AGC change remain to be further investigated.

Boreal forests transitioned from a carbon sink ($9.9\,\text{TgC\,yr}^{-1}$ and $85.8\,\text{TgC\,yr}^{-1}$ in the first two decades from 1988 to 2010) to a weak carbon source ($-2.5\text{TgC\,yr}^{-1}$) in 2011--2021 (Supplementary Table~\ref{tab:decadal_change_trend}).
Growing disturbance pressures, including fire, insect outbreaks, and logging~\cite{wangDisturbanceSuppressesAboveground2021} may have contributed to these substantial AGC losses.
Spatial patterns highlight marked declines across parts of eastern Eurasian boreal zones (Fig.~\ref{fig:fig4}), while some areas, such as western Siberia, show signs of recovery in 2011--2021 (Supplementary Table 3).
Canada's boreal forests exhibited strong decadal variability, alternating between source and sink over the three decades covered by our dataset (Supplementary Table 3), likely reflecting natural variability, possibly in combination with the influence of regional disturbances~\cite{chenContributionsInsectsDroughts2018}.

While temperate forests sequestered an average of $94.0\,\text{TgC\,yr}^{-1}$ and dry tropical \& subtropical forests stored about $59.2\,\text{TgC\,yr}^{-1}$ over 1988--2021, regions such as Europe and Australia show sink-to-source shifts accompanied by high net AGC losses. 
In Europe, forests have transitioned to a weak source of $-10.5\,\text{TgC\,yr}^{-1}$ in the last decade, when regarding the net AGC change, despite retaining a small positive stock trend of $2.6\,\text{TgC\,yr}^{-1}$. 
Previous studies attributed these losses to climate-related storms, pests (e.g., bark beetles), droughts, and fires~\cite{seidlIncreasingForestDisturbances2014}. 
Australian forests remained a net carbon source over the full period since 1988 ($-2.5\,\text{TgC\,yr}^{-1}$ on average), as short-term recovery was offset by major drought and fire events, particularly in 2019--2020~\cite{bowmanSeverityExtentAustralia2021}. 

Beyond these decadal sink-source shifts, we analyze interannual variability in AGC fluxes to identify which biomes drive year-to-year fluctuations in global forest AGC.
We adopt the flux partitioning approach developed by Ahlström et al.~\cite{ahlstromDominantRoleSemiarid2015}, which considers both the magnitude and the correlation of regional flux anomalies relative to the global signal.
We note that these estimates reflect variations solely in the AGC pool; total land-atmosphere carbon fluxes are additionally influenced by variability in belowground and soil carbon pools, which are not captured here.
We estimate that dry tropical and subtropical forests contributed 37\% of the interannual variability in global AGC fluxes over the past 30 years (1989--2021), followed by moist tropical forests (25\%), boreal forests (23\%), and temperate forests (15\%) (Supplementary Fig. 12). 
These fractions indicate that dry tropical \& subtropical terrestrial forest ecosystems are major drivers of the global interannual variability, consistent with earlier findings~\cite{ahlstromDominantRoleSemiarid2015}, and potentially linked to the vulnerability of these ecosystems to climatic impacts.
However, these contributions have changed over recent decades (Supplementary Fig. 12).
Temperate and boreal forests collectively dominate in the first (1989--2000) and third decade (2011--2021), accounting for 71\% and 90\% of the variability, respectively.
In contrast, during the second decade (2001--2010), moist tropical and dry tropical \& subtropical forests take the lead, contributing 86\%.
These temporal shifts may reflect changing disturbance regimes across regions.
For instance, warming-related stressors and insect outbreaks may have exerted a stronger influence in high-latitude forests during the first and third decade~\cite{wangDisturbanceSuppressesAboveground2021}, whereas tropical ecosystems in the second decade appear to have been more strongly affected by deforestation and degradation~\cite{lapolaDriversImpactsAmazon2023}.

\subsection*{Tropical AGC dynamics}\label{subsec23}

Given that our reconstruction reveals more pronounced decadal variability in tropical AGC compared to other regions (Fig.~\ref{fig:fig4} and Supplementary Table 3), we now focus specifically on the dynamics of these critical ecosystems.
To investigate tropical AGC dynamics and their coupling with the global carbon cycle, we analyze the correlation between interannual AGC fluxes and the atmospheric CO\textsubscript{2} growth rate, both linearly detrended over the 1988--2021 period.
By computing these correlations across distinct decadal intervals, we observe a strengthening negative relationship over recent decades, reaching $r=-0.63\,(p<0.05)$ in the most recent decade (Fig.~\ref{fig:fig5}a).
A 10‑year moving‑window analysis and uncertainty quantification via bootstrapping corroborate this increasingly negative correlation (Supplementary Fig. 13).
These observations may point to a strengthening role for tropical forest AGC in modulating the terrestrial carbon cycle variability. 

To understand spatial patterns within the tropics, we assess the relative contributions of different regions to overall AGC variability and examine the heterogeneity of AGC dynamics.
Again using the flux partitioning method~\cite{ahlstromDominantRoleSemiarid2015}, we find that the interannual variability in tropical AGC fluxes originates mainly from tropical America and tropical Africa (each contributing $\approx 46\%$), whereas tropical Asia accounts for only 7\% (see bars labelled 'All', representing the full period, in Fig.~\ref{fig:fig5}b). 
Within these continents, the Amazon rainforest dominates tropical America’s contribution by accounting for 69\% of its interannual variability, followed by the Congo rainforests contributing 24\% within tropical Africa, and the Indonesian rainforests contributing 6\% within tropical Asia.
These patterns underscore Amazon’s pivotal role in shaping tropical AGC dynamics at the interannual scale.

Although tropical forests gained a total of $1.3\,\text{PgC}$ from 1988 to 2021, with a trend of $2.7\,\text{TgC\,yr}^{-1}$, this overall balance masks pronounced regional disparities (Supplementary Table 3).
We find that neither tropical American forests overall nor the Amazon rainforest in particular have fully returned to their 2003 AGC levels, although partial recovery occurred after several subsequent AGC losses (Fig.~\ref{fig:fig5}d,e).
In contrast, AGC in tropical African forests and the Congo rainforest in particular show a relatively stable period between 1998 and 2011, and a substantial decrease in 2015/16.
Since then, their AGC trajectories have diverged: Tropical African forests have largely recovered, approaching the 2014 peak, whereas the Congo Basin has remained $\sim$0.3 PgC below its 2014 level (Fig.~\ref{fig:fig5}f,g).
In tropical Asia, the divergence in AGC dynamics between continental-scale forests and the Indonesian rainforest is particularly pronounced.
Tropical Asian forests exhibited an overall upward trend ($2.2\,\text{TgC\,yr}^{-1}$), reaching a peak in 2015 followed by a sharp decline in 2016, with partial recovery in subsequent years that has not returned to the peak level. 
Meanwhile, the Indonesian rainforests followed a long-term declining trajectory ($-3.6\,\text{TgC\,yr}^{-1}$)(Fig.~\ref{fig:fig5}h,i and Supplementary Table 3).
The continued deforestation and land-use change in the Congo and Indonesian rainforests likely contributed to their divergence from continental AGC trends~\cite{liDeforestationinducedClimateChange2022,wangHighresolutionMapsShow2023}.

Previous studies have reported that certain major tropical drought events were associated with substantial carbon losses~\cite{liuContrastingCarbonCycle2017,wigneronTropicalForestsDid2020}.
To examine such dynamics in our dataset, we compare regional AGC dynamics with moisture availability conditions quantified as the fraction of grid cells with three-month Standardised Precipitation Evapotranspiration Index (SPEI3) $\leq -1$~\cite{tirivaromboDroughtMonitoringAnalysis2018} (see background colouring in Fig.~\ref{fig:fig5}c-i).
While localized AGC losses co-occur with severe moisture stress during specific events, the broader regional relationship is often less distinct.
For example, the Congo and Indonesia rainforests occasionally exhibit sharp declines in AGC without corresponding peaks in drought fraction (Fig.~\ref{fig:fig5}g,i).
Such mismatches likely reflect confounding influences, including concurrent anthropogenic disturbances and other climate extremes, that also shape AGC dynamics.
However, for the Amazon, we specifically examine four well-documented large-scale drought events (1997/98~\cite{siegertIncreasedDamageFires2001}, 2004/05~\cite{aragaoSpatialPatternsFire2007}, 2009/10~\cite{lewis2010AmazonDrought2011}, and 2015/16~\cite{bennettSensitivitySouthAmerican2023}) and indeed observe pronounced minima in Amazon AGC stock during the latter three drought years (Fig.~\ref{fig:fig5}e).
This alignment suggests that climate extremes are driving substantial carbon losses in the Amazon.

\subsection*{AGC loss in the Brazilian Amazon under compound disturbances}\label{subsec24}

The interannual variability in AGC fluxes across tropical forests stems from a complex interplay between human activities, climatic variability, and plant physiological responses~\cite{gattiAmazoniaCarbonSource2021}.
Focusing on the Brazilian Amazon as a critical hotspot of this variability~\cite{lovejoyAmazonTippingPoint2018,floresCriticalTransitionsAmazon2024}, we investigate the drivers of gross AGC losses (defined as the spatiotemporal aggregation of AGC decreases, excluding gains; see Methods). 
To achieve this, we integrate long-term deforestation records from PRODES~\cite{institutonacionaldepesquisasespaciaisinpePRODESCoordenacaoGeralObservacao2025} with the Intact Forest Landscapes (IFL) dataset~\cite{potapov2017last}. 
This allows us to distinguish between intact forests and deforested regions (Methods).
In intact forests, gross AGC losses are likely driven by natural factors (e.g., droughts, fires, storms, and tree mortality) and indirect anthropogenic impacts (e.g., deforestation-induced edge effects and precipitation alterations)~\cite{lapolaDriversImpactsAmazon2023,qinImpactAmazonianDeforestation2025}. 
In contrast, gross AGC losses in deforested regions are more strongly associated with land-use changes, including logging, agricultural expansion, and infrastructure development.

Our partitioning analysis of the interannual variability of gross AGC losses reveals changing contributions from human- and nature-related factors over three decades (Fig.~\ref{fig:fig6}).
During 1988--2000, AGC losses in deforested areas accounted for 60\% of the interannual variability in total gross losses, while intact forests contributed 33\%.
In the second decade (2001--2010), this balance changed markedly: the contribution of deforested regions dropped to 32\%, whereas that of intact forests surged to 59\%.
This shift might reflect, in part, the influence of stricter deforestation-reduction policies introduced after 2004 (Fig.~\ref{fig:fig6}a), which suppressed the variability from direct clearing, alongside severe climate anomalies driving fluctuations in intact forests.
In the final decade (2011--2021), although deforestation rates began to increase again after 2012 (Fig.~\ref{fig:fig6}a), the average annual deforested area (7,681 km$^2$ yr$^{-1}$) remained lower than in the preceding two decades (16,959 km$^2$ yr$^{-1}$ during 1988--2000 and 16,531 km$^2$ yr$^{-1}$ during 2001--2010).
Consequently, this persistently lower baseline of direct clearing reduced its influence on year-to-year fluctuations, with deforested areas contributing only 13\% to interannual variability. 
In contrast, intact forests dominated the variability (76\%), indicating a growing role of other disturbances, and potential environmental stress induced by anthropogenic climate change on carbon loss in intact forests.

Spatially explicit maps of decadal gross AGC losses also corroborate these patterns (Fig.~\ref{fig:fig6}f-h).
From 1988 to 2000, high gross losses were heavily concentrated along the so-called “Arc of Deforestation” (Fig.~\ref{fig:fig6}f), implicating direct human clearing as the principal driver.
However, in 2001--2010, this loss footprint expanded extensively across both deforested regions and intact forests, probably reflecting a more widespread effect of disturbances.
By 2011--2021, the primary losses occurred in intact forests, and the deforestation arc signature weakens in the gross-loss maps (a pattern also evident in Fig.~\ref{fig:fig3}d and Fig.~\ref{fig:fig4}d).
These spatial patterns are closely associated with the evolving spatial distribution of the deforestation fraction (Fig.~\ref{fig:fig6}d,e) and align geographically with key degradation drivers identified in previous research~\cite{lapolaDriversImpactsAmazon2023}.

\section*{Discussion}\label{sec3}

Understanding long-term AGC dynamics in natural forests is vital for assessing their carbon sequestration potential under anthropogenic forcing. 
However, reconstructing these dynamics presents ongoing challenges, largely constrained by the limited availability of historical ``ground truth'' observations~\cite{yangGlobalIncreaseBiomass2023}.
Our analysis indicates that current observational AGC records naturally entail uncertainties that are often sensitive to the choice of input data and methodological frameworks.
Even products derived from comparable input data (e.g., L-band VOD) and empirical approaches may still yield varying estimates of temporal AGC variability and long-term trends (e.g., AGC products by Fan et al.~\cite{fanSatelliteobservedPantropicalCarbon2019} and Boitard et al.~\cite{boitardAbovegroundBiomassDataset2025}). 
Our empirical findings thus underscore the value of considering these structural nuances when interpreting short-term, grid-cell-level AGC fluxes.

The recent availability of bias-corrected, state-of-the-art Earth observation records, encompassing diverse vegetation variables and environmental data~\cite{zottaVODCAV2Multisensor2024,liSpatiotemporallyConsistentGlobal2023,caoSpatiotemporallyConsistentGlobal2023,harper29yearTimeSeries2023,wangNewSetMODIS2020,macferrinEarthTopography20222024}, provides complementary information to reduce these structural uncertainties. 
Building on this opportunity, we developed a probabilistic deep learning method to reconstruct a long-term, internally consistent AGC record, accompanied by explicit uncertainty estimates.
The spatially resolved SNR analysis underpins the reliability of our estimated AGC fluxes and trends (Supplementary Figs. 10 and 11 and Supplementary Table 4). 
These evaluations also confirm that aggregating short-term fluxes over regional and longer decadal scales substantially suppresses local noise, yielding robust signals.

\begin{figure}[H]
\centering
\includegraphics[width=1\linewidth]{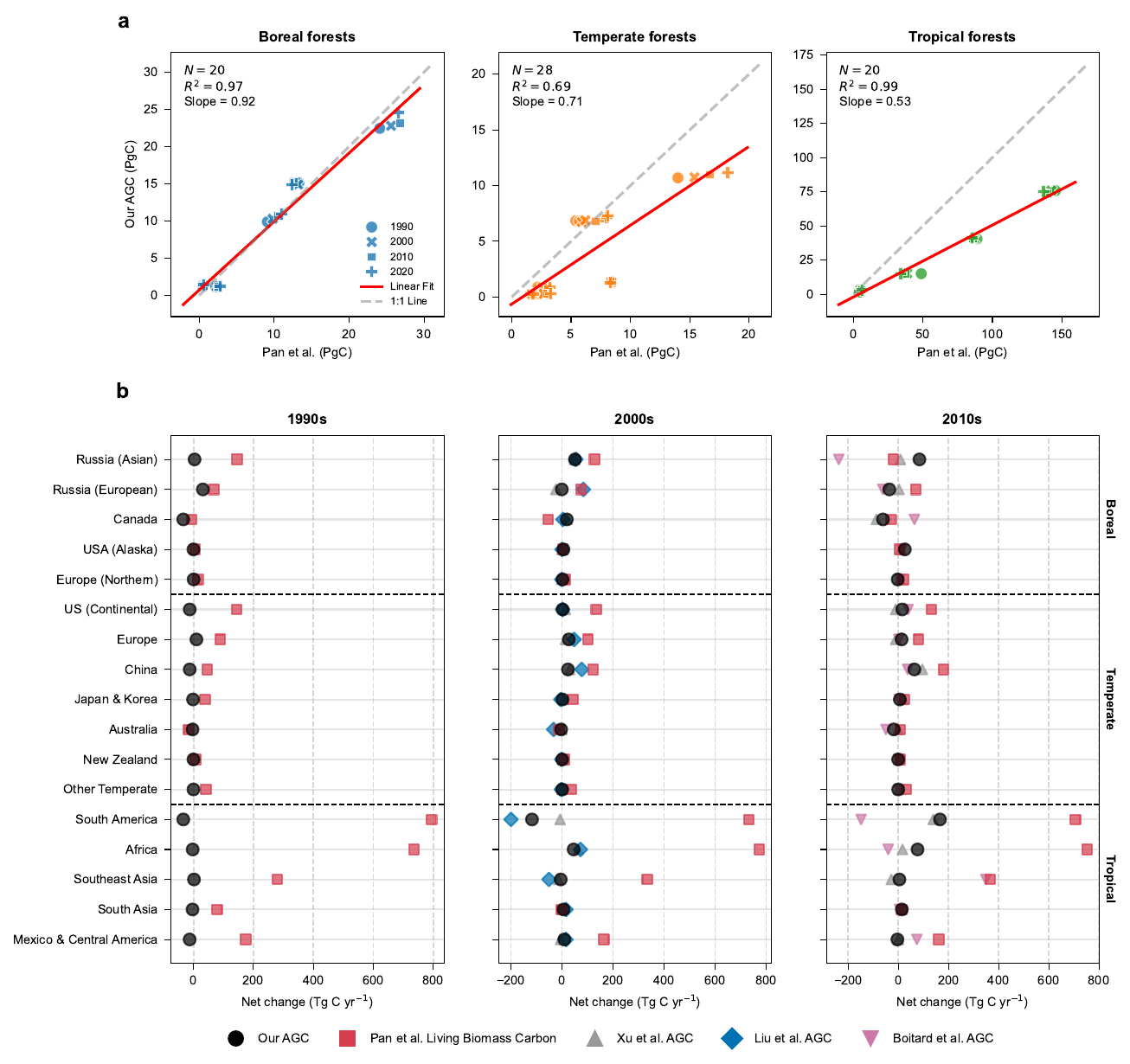}
\caption{
\textbf{Comparison of our forest AGC estimates with national inventory-based data and other remote sensing products.} 
\textbf{a}, Scatter plots comparing our regional AGC stocks against the living biomass carbon stocks reported by Pan et al.~\cite{panEnduringWorldForest2024} across boreal, temperate, and tropical forests. Each data point represents a specific geographic region or country for a given year (1990, 2000, 2010, or 2020), following the definitions in Pan et al.~\cite{panEnduringWorldForest2024}. Solid red lines indicate the linear fits, and dashed grey lines represent the 1:1 relationships. The consistent slopes of $<1$ reflect the fundamental distinction between our estimator (AGC only) and Pan et al.'s estimator (total living biomass carbon, which includes below-ground components). 
\textbf{b}, Comparison of decadal net carbon stock changes across regions for the periods 1990s (1990--1999), 2000s (2000--2009), and 2010s (2010--2019). Our AGC change estimates (black circles) are compared alongside inventory-based living biomass changes (red squares) by Pan et al.~\cite{panEnduringWorldForest2024} and other available remote sensing-based AGC products (Xu et al.~\cite{xuChangesGlobalTerrestrial2021}, Liu et al.~\cite{liuRecentReversalLoss2015}, and Boitard et al.~\cite{boitardAbovegroundBiomassDataset2025}). Regions are vertically grouped by their corresponding dominant biomes following the definitions in Pan et al.~\cite{panEnduringWorldForest2024}. 
}
\label{fig:fig1}
\end{figure}

\begin{figure}[H]
\centering
\includegraphics[width=1\linewidth]{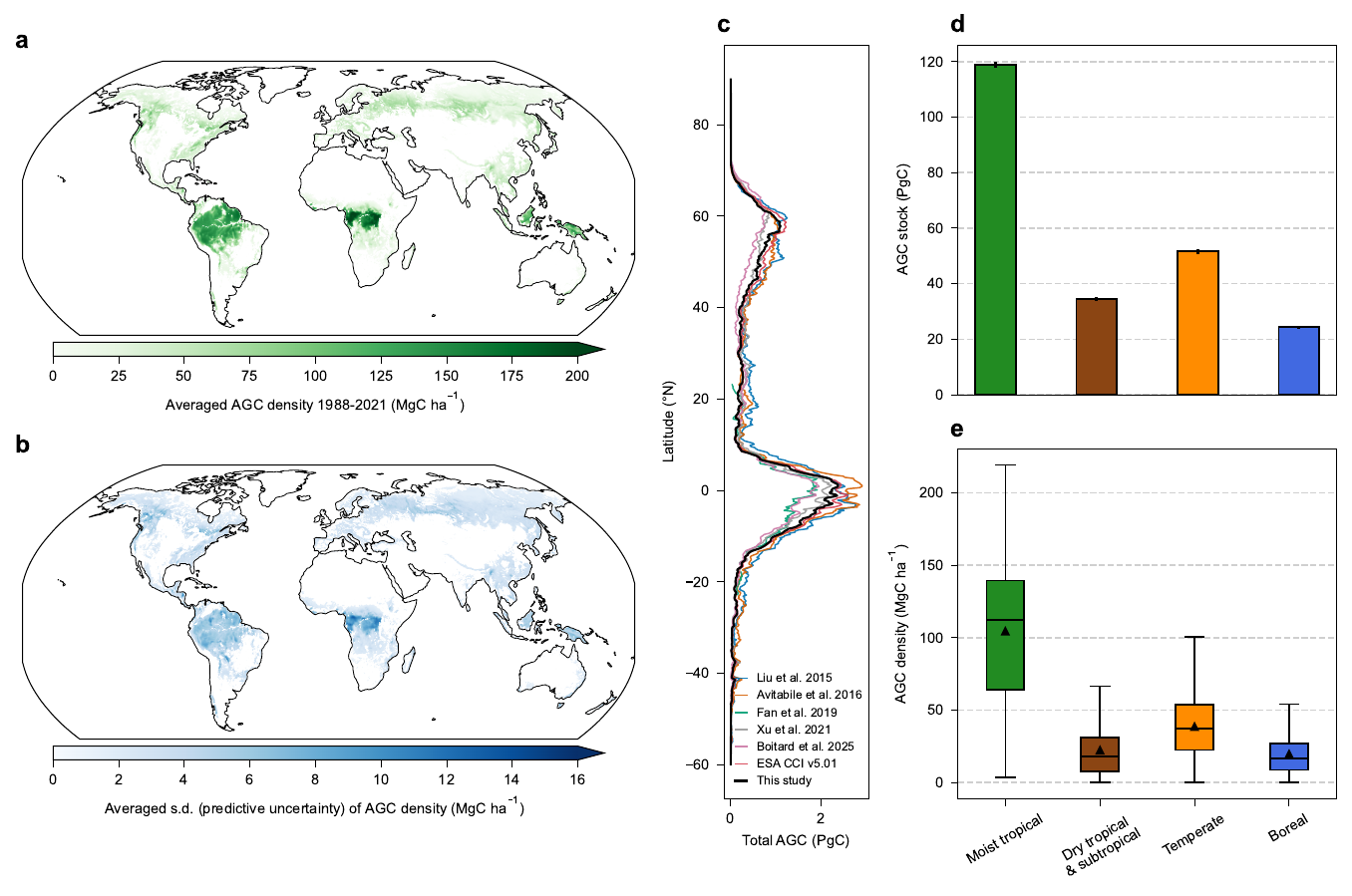}
\caption{
\textbf{Spatial patterns, predictive uncertainty, and biome-level distribution of global forest AGC.} 
\textbf{a}, Multi-year averaged forest AGC density at 0.25\textdegree{} resolution, calculated from the long-term (1988--2021) time series reconstructed in this study. 
\textbf{b}, Multi-year averaged predictive uncertainty (standard deviation), which jointly reflects observation noise and model underrepresentation (see Methods).
\textbf{c}, Latitudinal profiles of zonal AGC stock sums. Profiles represent multi-year means for each reference dataset, except for ESA CCI AGC, which is averaged over 2010 and 2021 to exclude our model training period.
\textbf{d}, Total AGC stocks of different biomes, with moist tropical forests dominating (119 PgC, 52\%), followed by temperate (52 PgC, 23\%), dry tropical \& subtropical (35 PgC, 15\%), and boreal (24 PgC, 11\%) forests. Black error bars indicate the uncertainty range.
\textbf{e}, Boxplots of biome-specific AGC density at the grid-cell level, exhibiting a similar cross-biome gradient to the total stocks. The boxplot boundaries from top to bottom represent the maximum, third quartile, median, first quartile, and minimum, and black triangles mark the mean.
}
\label{fig:fig2}
\end{figure}

\begin{figure}[H]
\centering
\includegraphics[width=1\linewidth]{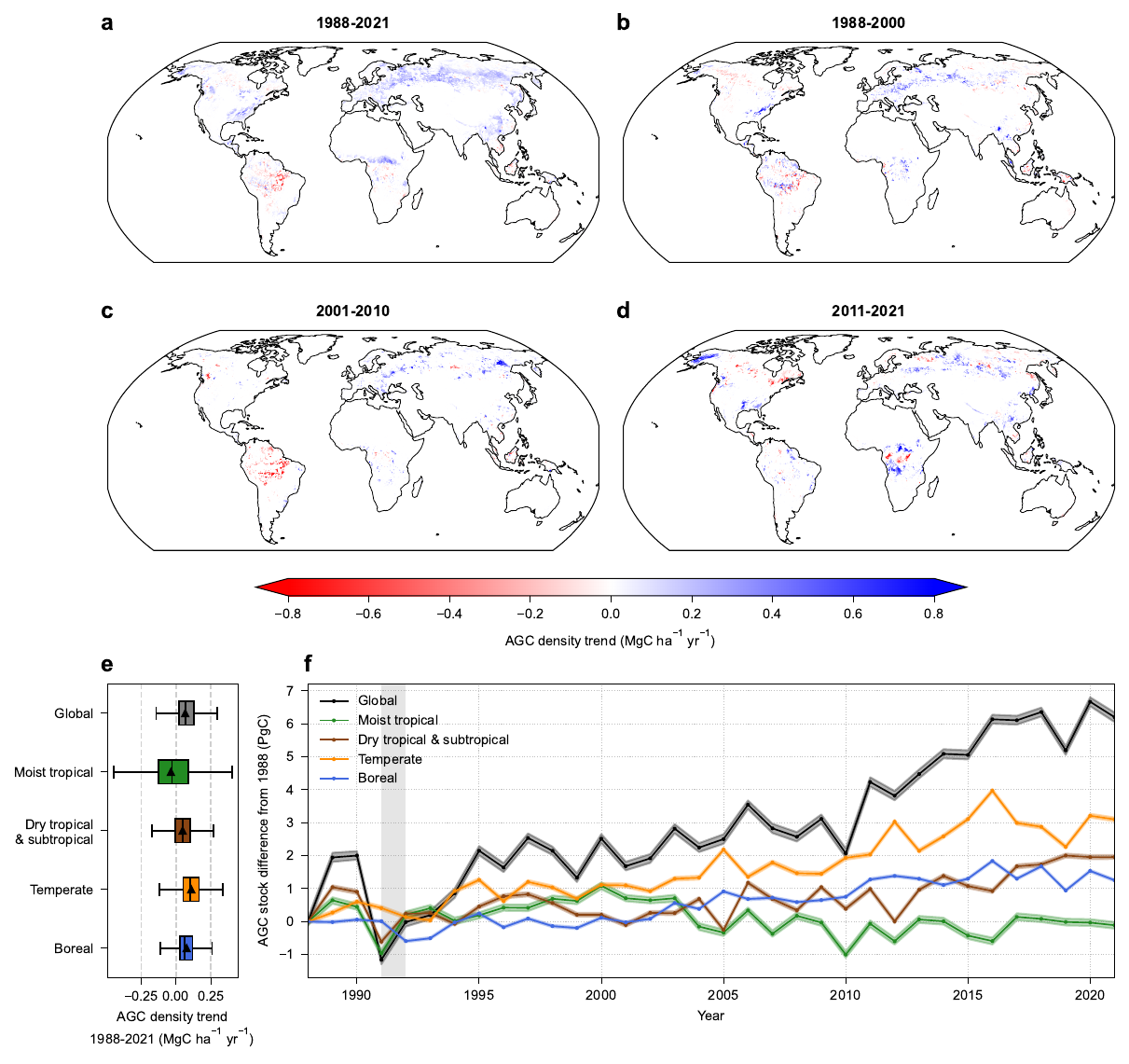}
\caption{
\textbf{Spatially explicit trends and biome-level dynamics of global forest AGC.}
\textbf{a}, Overall AGC density trends across the entire study period at 0.25\textdegree{} resolution.
\textbf{b}--\textbf{d}, Decadal AGC density trends for the periods 1988--2000 (\textbf{b}), 2001--2010 (\textbf{c}), and 2011--2021 (\textbf{d}). Grid-cell-wise trends are computed via the Theil–Sen slope and a modified Mann–Kendall test to account for serial autocorrelation, with increases shown in blue and declines in red, retaining only grid cells with $p < 0.05$.
\textbf{e}, Boxplots of AGC density trends for the globe (grey) as well as moist tropical (green), dry tropical \& subtropical (brown), temperate (orange), and boreal (blue) forests. Each box denotes the median, quartiles, and range; black triangles indicate mean values. 
\textbf{f}, Time series of AGC stock changes with respect to 1988 at both global and biome levels, revealing overall increasing AGC stocks in global, temperate, boreal, and dry tropical \& subtropical forests, and slightly decreasing AGC in moist tropical forests. Only grid cells with valid data across all years are considered. Shaded areas depict the 95\% uncertainty interval (see Methods). The vertical grey band denotes the time period affected by the Mt. Pinatubo eruption (1991-1992), which is not included in the time series analysis.
}
\label{fig:fig3}
\end{figure}

\begin{figure}[H]
\centering
\includegraphics[width=1\linewidth]{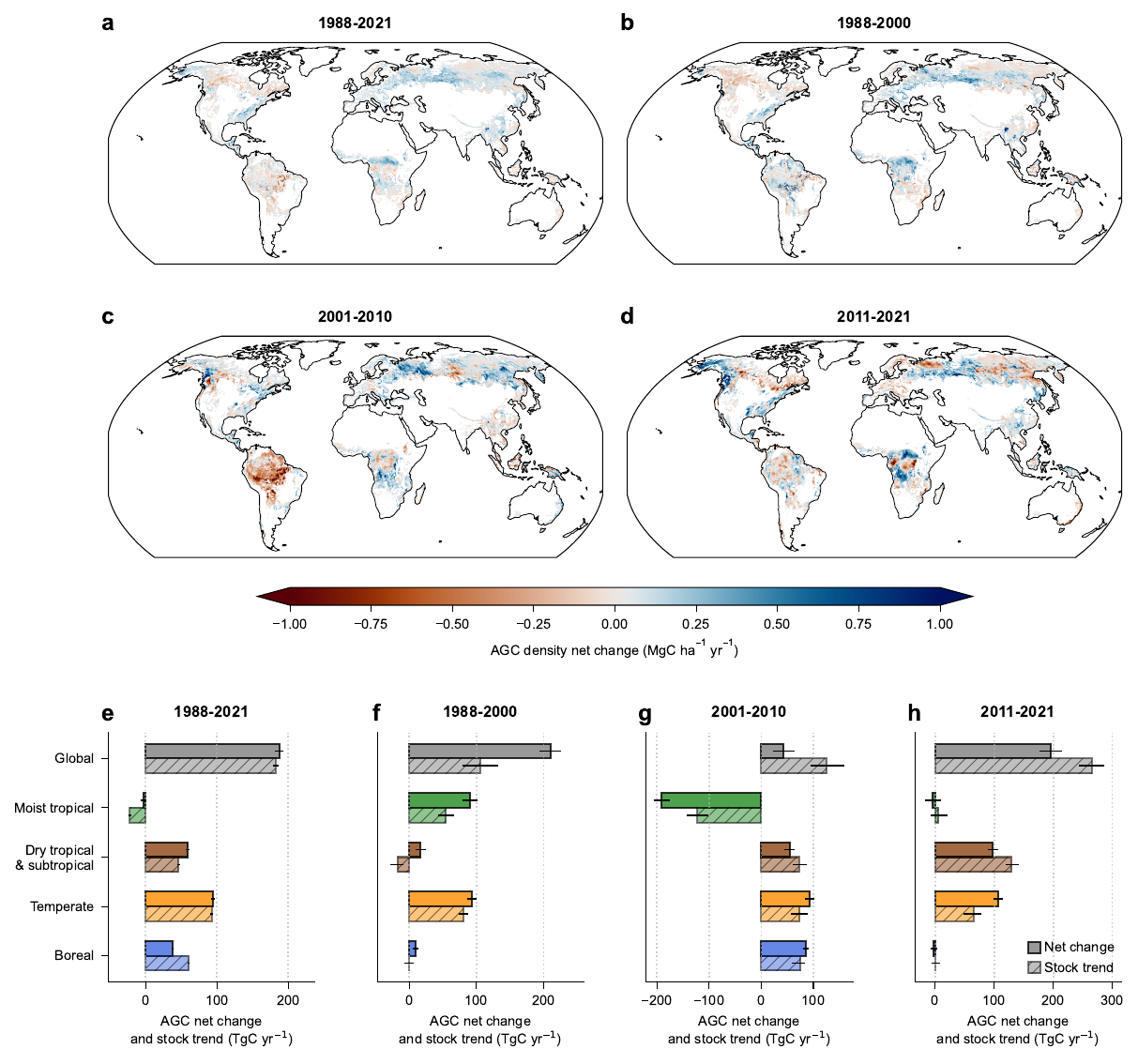}
\caption{ 
\textbf{Spatially explicit net AGC changes across decades from 1988 to 2021.}
\textbf{a}, Annual mean values of net AGC change over the full 1988--2021 period at 0.25\textdegree{} resolution. 
\textbf{b--d}, Annual mean values of net AGC changes for 1988--2000, 2001--2010, and 2011--2021, respectively. 
Positive changes (blue) indicate net AGC gains, whereas negative changes (brown) denote net AGC losses.
\textbf{e--h}, Global and biome-specific net changes and trends in AGC stocks for the corresponding time periods.
Net changes represent the annual mean AGC difference over each period, while trends are derived from Theil–Sen slope estimates of the AGC stock time series over the same corresponding period.
For trend calculation, we consider only grid cells with valid data across all years, and AGC stock values for 1991 and 1992 are ignored.
Error bars indicate the 95\% uncertainty range (see Methods).
}
\label{fig:fig4}
\end{figure}

\begin{figure}[H]
\centering
\includegraphics[width=1\linewidth]{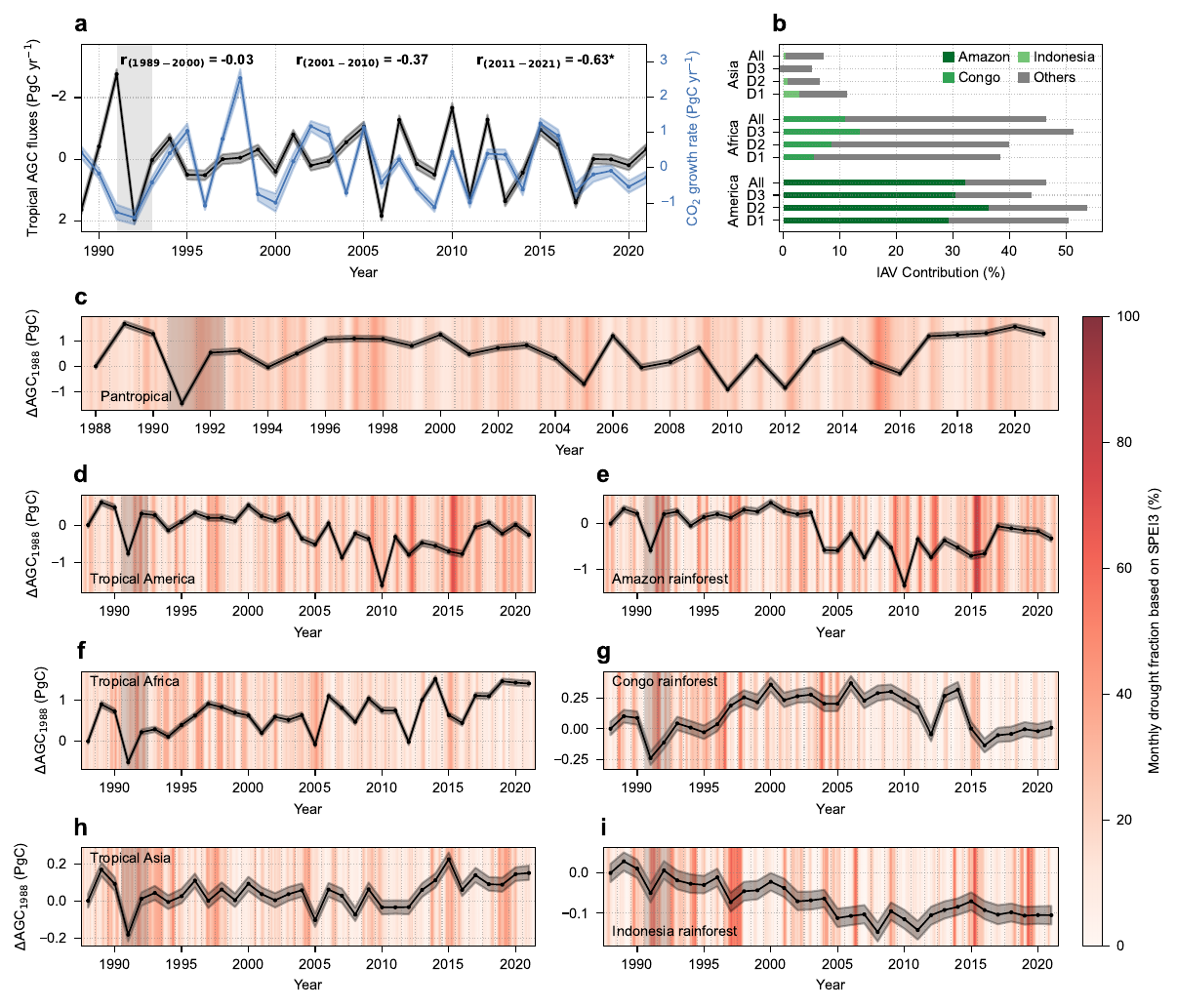}
\caption{
\textbf{Tropical AGC fluxes and stock changes over time.}
\textbf{a}, Annual AGC flux (black) over pan-tropical forests, spanning approximately 23.5\textdegree{}N to 23.5\textdegree{}S, compared with atmospheric CO\textsubscript{2} growth rate (blue), both with long-term linear trends removed to highlight interannual variability (IAV).
Asterisks (*) indicate statistical significance at $p < 0.05$, using the two-tailed t-test.
\textbf{b}, IAV contribution of different sub-regions to the overall interannual variability of pan-tropical AGC fluxes across tropical America, Africa, Asia during four time periods (D1: 1989--2000, D2: 2001--2010, D3: 2011--2021, All: 1989--2021). Sub-regions include the Amazon, Congo, and Indonesian rainforests, and the remaining non-rainforest areas.
\textbf{c}, Time series of AGC stock changes with respect to 1988 for all pan-tropical forests.
\textbf{d}--\textbf{i}, Corresponding AGC stock changes for tropical America (\textbf{d}), Amazon rainforests (\textbf{e}), tropical Africa (\textbf{f}), Congo rainforests (\textbf{g}), tropical Asia (\textbf{h}), and Indonesian rainforests (\textbf{i}). 
Only grid cells with valid data across all years are considered. Grey-shaded regions in each panel represent the 95\% uncertainty intervals (see Methods). Blue-shaded regions represent the uncertainties in CO\textsubscript{2} growth rates.
The vertical grey band is the period of the Mt. Pinatubo eruption, not included in the time series analysis.
Vertical red shading represents the drought-affected area fraction across tropical forest grid cells, based on the three-month Standardized Precipitation Evapotranspiration Index (SPEI3). 
Grid cells with SPEI3$\leq -1$ are classified as drought-affected.
}
\label{fig:fig5}
\end{figure}

\begin{figure}[H]
\centering
\includegraphics[width=1\linewidth]{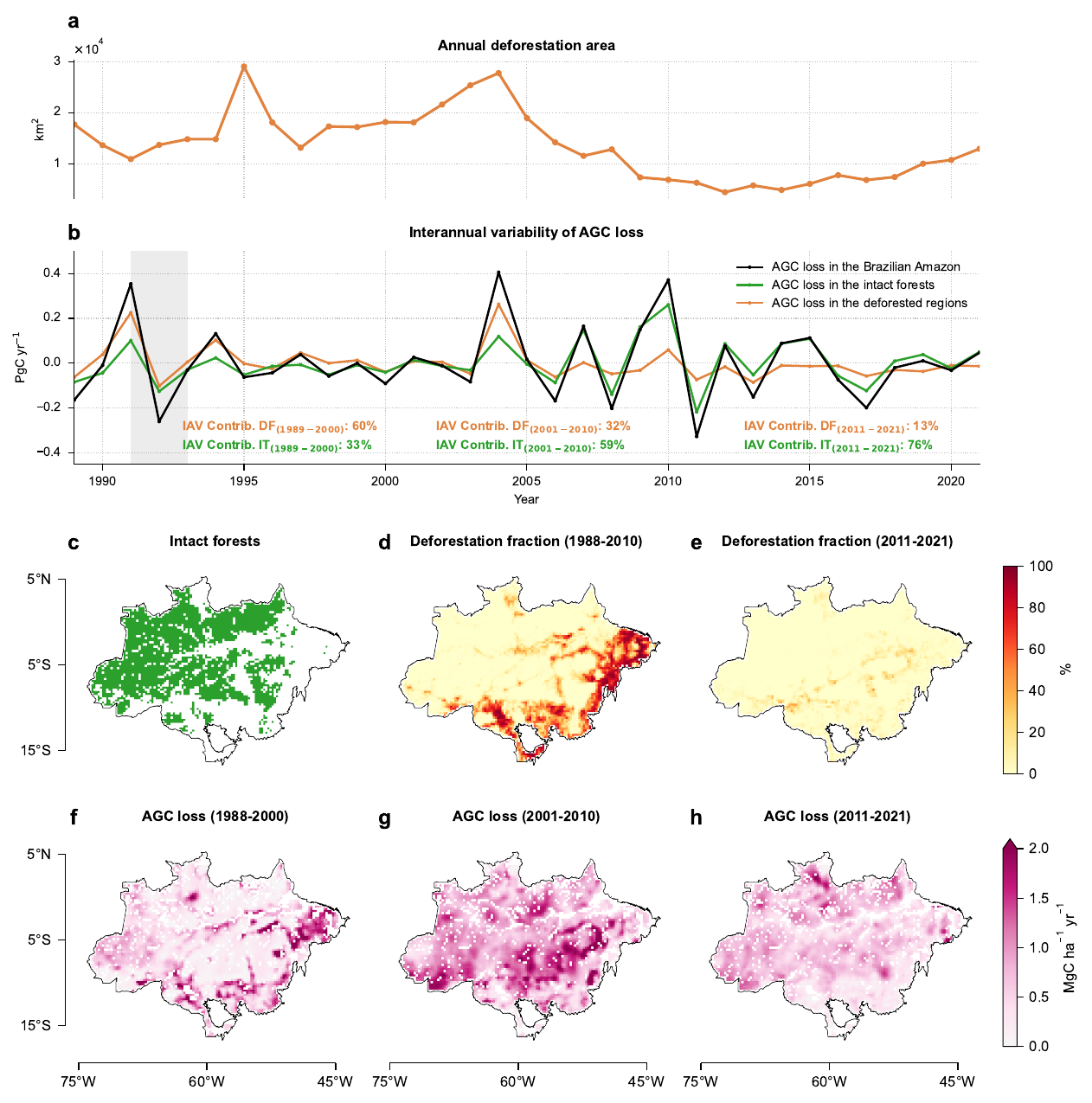}
\caption{
\textbf{Spatiotemporal dynamics of AGC losses in the Brazilian Amazon.}
\textbf{a}, Time series of annual deforestation area, as reported by the Brazilian National Institute for Space Research (INPE)~\cite{institutonacionaldepesquisasespaciaisinpePRODESCoordenacaoGeralObservacao2025}.
\textbf{b}, Interannual variability (IAV) of regional AGC gross loss fluxes, partitioned into contributions from deforested regions (orange) and intact forests (green). The grey shaded band indicates the period affected by the Mt. Pinatubo eruption, which is excluded from the temporal analysis.
\textbf{c}, Spatial distribution mask of intact forests (green), derived from the Intact Forest Landscapes (IFL) project~\cite{potapov2017last}.
\textbf{d}, \textbf{e}, Spatial patterns of deforestation fraction aggregated at 0.25\textdegree{} resolution for 1988--2010 (\textbf{d}) and 2011--2021 (\textbf{e}) from INPE PRODES~\cite{institutonacionaldepesquisasespaciaisinpePRODESCoordenacaoGeralObservacao2025}.
\textbf{f}--\textbf{h}, Spatially explicit patterns of annual AGC gross loss at 0.25\textdegree{} resolution for three decadal intervals: 1988--2000 (\textbf{f}), 2001--2010 (\textbf{g}), and 2011--2021 (\textbf{h}).
}
\label{fig:fig6}
\end{figure}

While cross-comparisons among existing short-term reference datasets reveal discrepancies, our reconstruction demonstrates generally higher correlations with these diverse datasets (Table~\ref{tab:global_agc_r} and Supplementary Note 1).
The aggregated long-term AGC stocks and changes derived from our reconstruction generally align with recent forest inventory assessments~\cite{panEnduringWorldForest2024}, despite the persistent methodological disparities between continuous remote sensing observations and sparse plot networks, particularly in tropical regions (Fig.~\ref{fig:fig1}). 
Our spatially explicit reconstruction thus serves as a critical complement to both recent inventory-based global reports~\cite{panEnduringWorldForest2024} and multi-source harmonized estimates~\cite{bar-onRecentGainsGlobal2025} (see Supplementary Discussions).
By extending the temporal coverage of reliable global AGC estimates by 10--20 years, our reconstructed record enables a more robust assessment of forest carbon dynamics over the past three decades.

While global forests broadly remain a net AGC sink, our reconstructed multi-decadal record reveals noteworthy sink-to-source transitions in moist tropical and boreal biomes, alongside distinct decadal shifts in the regions driving interannual carbon fluctuations~\cite{ahlstromDominantRoleSemiarid2015}.
Crucially, our analysis indicates a strengthening negative coupling between tropical AGC fluxes and the atmospheric CO\textsubscript{2} growth rate over the past three decades. 
Building on prior hypotheses that attributed the intensifying water-carbon coupling primarily to ENSO-induced water stress~\cite{liuIncreasinglyNegativeTropical2023}, we investigate these climatic drivers across our 34-year continuous record (see Supplementary Discussion).
While our dataset captures the severe AGC losses associated with the extreme 2015/16 El Niño event, we found insufficient evidence to attribute the overall large-scale interannual carbon anomalies to ENSO phases (Supplementary Fig. 17).
The absence of a widespread correlation may partly indicate the inherent uncertainties in AGC reconstructions, which may hinder the full detection of ENSO impacts on AGC dynamics.

As a critical hotspot of tropical carbon dynamics, the Brazilian Amazon shows vulnerability to compound disturbances.
Over the past three decades, the regional dynamics driving the interannual variability of gross AGC losses have undergone a structural transition: the contribution of deforested areas decreased from 60\% to 13\%, whereas that of intact forests increased from 33\% to 76\%. 
This multi-decadal trajectory contextualizes shorter-term degradation patterns, indicating a shift in carbon emission drivers from direct land clearing to climate-induced tree mortality and indirect edge effects~\cite{boersDeforestationinducedTippingPoint2017,qinCarbonLossForest2021,bochowSouthAmericanMonsoon2023}. 
The dominating role of carbon losses from intact forests underscores a critical susceptibility to anthropogenic climate change~\cite{mitchardTropicalForestCarbon2018}. 
Consequently, global carbon management frameworks, such as the Paris Agreement and Reducing Emissions from Deforestation and Forest Degradation (REDD+) initiatives, must urgently evolve to reflect these shifting drivers.
Moving beyond merely curbing clear-cutting, national and international strategies should be operationalized to prioritize the protection and resilience of intact forest ecosystems against accumulating climate stress.
While fully translating these insights into actionable policies requires integrating complex socioecological dimensions and biophysical climate feedbacks, an endeavor that falls beyond the scope of our study, we hope our multi-decadal baseline will actively facilitate these crucial future efforts.

Several methodological limitations should be noted when interpreting our reconstructed AGC dynamics.
First, our analysis specifically focuses on global forest AGC, not total biomass carbon.
Total biomass carbon, including both above- and below-ground components, is commonly estimated using biome-specific root-to-shoot ratios~\cite{xuChangesGlobalTerrestrial2021,yangGlobalIncreaseBiomass2023} or empirical functions~\cite{fengDoublingAnnualForest2022}.
These approaches introduce additional uncertainty into carbon flux estimates, particularly since the interannual variability in the relationship between AGB and below-ground biomass remains poorly understood, and responses of AGB and below-ground biomass to climate and anthropogenic disturbances can also differ~\cite{zhangGlobalSynthesisBelowground2015,spawnHarmonizedGlobalMaps2020}.
While a formal causal attribution of AGC changes is beyond the scope of this study, we interpret observed AGC dynamics in the context of existing literature and link them to both climatic and anthropogenic influences.
These region-specific interpretations regarding the drivers of carbon loss are further corroborated by spatially explicit classifications of global forest disturbance drivers~\cite{curtisClassifyingDriversGlobal2018} (Supplementary Fig. 15).

We acknowledge that persistent gaps and sensor transitions in early satellite observations could introduce regional uncertainties or potential underestimations of historical AGC stocks.
Due to the inherent absence of a continuous, fine temporal-resolution ground truth, relying on space-for-time substitution is an unavoidable and widely established standard practice within the remote sensing community for AGC reconstructions~\cite{liuRecentReversalLoss2015,brandtSatellitePassiveMicrowaves2018,fanSatelliteobservedPantropicalCarbon2019,qinCarbonLossForest2021,yangGlobalIncreaseBiomass2023,boitardAbovegroundBiomassDataset2025}.
However, this approach inherently assumes that spatial predictor-AGC relationships remain stable over time, a simplification that may not fully capture the evolving impacts of climate and land-use changes.
To account for this, our probabilistic framework explicitly generates predictive uncertainty bounds.
These estimated uncertainty intervals do not exhibit notable widening in earlier periods,  suggesting that the predictor distributions remained largely consistent and did not shift out of distribution drastically over the 34-year period.
Moreover, because our training period captures broad recent climate variability, climate conditions over the past 30 years generally fall within this observed range, making historical reconstruction more reliable than projecting unknown future AGC dynamics.

Despite the inherent constraints of space-for-time substitution and the probabilistic Gaussian assumptions, our uncertainty-aware AGC record provides an internally consistent reference for tracking terrestrial AGC trajectories.
Future studies explicitly quantifying the impacts of extreme climatic events and human disturbances using our AGC estimates could hence provide valuable insights into the evolving drivers of global forest carbon dynamics.

\clearpage

\clearpage

\clearpage

\clearpage

\clearpage

\clearpage

\clearpage
\begin{table}[htbp]
    \centering
    \caption{Comparison of model performance in estimating AGC.}
    \begin{tabular}{l|cc|c}
        \toprule
        \multirow{2}{*}{\textbf{Method}} & \multicolumn{2}{c|}{\textbf{Grid-cell-wise comparison}} & \textbf{Global AGC time series} \\
        \cmidrule(lr){2-3} \cmidrule(lr){4-4}
        & R\textsuperscript{2} & RMSE ($\mathrm{MgC\,ha^{-1}}$) & Pearson's $r$ \\
        \midrule
        \multicolumn{4}{l}{\textit{VOD-only empirical models}} \\
        \midrule
        Arctan (CXKu-band)  & 0.22 & 70.26 & 0.35  \\
        Logistic (CXKu-band) & 0.21 & 70.78 & 0.35  \\
        Arctan (L-band)  & 0.64 & 47.18 & /  \\
        Logistic (L-band) & 0.35 & 63.33 & /  \\
        \midrule
        \multicolumn{4}{l}{\textit{Multivariate machine learning models}} \\
        \midrule
        Ridge regression    & 0.81 & 34.60 & 0.64** \\
        Lasso regression    & 0.78 & 37.72 & 0.62** \\
        Random forest       & 0.98 & 10.44 & 0.45* \\
        Our CNN    & 0.97 & 12.64 & 0.70*** \\
        \bottomrule
    \end{tabular}
    
    \vspace{1.5ex}
    {\footnotesize Baseline models (Arctan~\cite{liuRecentReversalLoss2015}, Logistic~\cite{boitardAbovegroundBiomassDataset2025}) use only VOD, whereas machine learning models use multivariate inputs. Grid-cell-wise evaluation is against the held-out ESA CCI AGB maps (2010, 2021). Temporal performance is assessed by calculating the Pearson correlation ($r$) between the estimated time series of global annual total AGC and the independent reference global totals from Xu et al.~\cite{xuChangesGlobalTerrestrial2021} (2000--2019). Temporal comparison for L-band VOD is omitted due to its limited data span (available since 2010). Significance (two-tailed t-test): * $p < 0.05$, ** $p < 0.01$, *** $p < 0.001$.}
    \label{tab:model_comparison}
\end{table}

\clearpage
\begin{table}[htbp]
    \centering
    \caption{Correlation of AGB time series among observational products across global biomes during overlapping years.}
    \label{tab:global_agc_r}
    
    \small
    \setlength{\tabcolsep}{6pt}
    
    \begin{tabular}{l c c c c c}
        \toprule
        \textbf{Comparison} & \textbf{Global} & \makecell{\textbf{Moist}\\\textbf{tropical}} & \makecell{\textbf{Dry trop. \&}\\\textbf{subtrop.}} & \textbf{Temperate} & \textbf{Boreal} \\
        \midrule
        \multicolumn{6}{l}{\textbf{Pearson's r for annual AGC stock time series}} \\
        \midrule
        \makecell[l]{Ours vs Liu et al.~\cite{liuRecentReversalLoss2015}\\(1993--2012)}         & -0.36   & 0.64** & 0.25   & 0.70*** & 0.61** \\
        \makecell[l]{Ours vs Xu et al.\cite{xuChangesGlobalTerrestrial2021}\\(2000--2019)}           & 0.70*** & 0.43   & 0.85***& 0.92*** & 0.59** \\
        \makecell[l]{Ours vs Boitard et al.\cite{boitardAbovegroundBiomassDataset2025}\\(2011--2021)} & -0.01   & -0.37  & 0.84** & 0.18    & 0.19   \\
        \makecell[l]{Liu et al. vs Xu et al.\\(2000--2012)}                  & -0.59* & 0.36   & 0.27   & 0.53    & 0.30   \\
        \makecell[l]{Xu et al. vs Boitard et al.\\(2011--2019)}              & 0.27    & -0.65  & 0.84** & 0.47    & -0.18  \\
        \midrule
        \multicolumn{6}{l}{\textbf{Pearson's r for annual AGC flux time series}} \\
        \midrule
        \makecell[l]{Ours vs Liu et al.\\(1994--2012)}                       & 0.07    & 0.54* & 0.47* & 0.08    & -0.20  \\
        \makecell[l]{Ours vs Xu et al.\\(2001--2019)}                        & 0.18    & 0.49* & 0.53* & 0.83*** & 0.20   \\
        \makecell[l]{Ours vs Boitard et al.\\(2012--2021)}                   & 0.27    & -0.33  & 0.27   & 0.00    & 0.25   \\
        \makecell[l]{Liu et al. vs Xu et al.\\(2001--2012)}                  & -0.23   & 0.12   & 0.23   & 0.26    & 0.06   \\
        \makecell[l]{Xu et al. vs Boitard et al.\\(2012--2019)}              & 0.13    & -0.60  & -0.22  & -0.10   & -0.27  \\
        \bottomrule
    \end{tabular}
    
    \footnotetext{Significance (two-tailed t-test): * $p < 0.05$, ** $p < 0.01$, *** $p < 0.001$.}
\end{table}

\clearpage
\section*{Methods}\label{sec4}

\subsection*{Modeling predictors}\label{predictors}

We use multi-source vegetation variables and environmental data to reconstruct spatially explicit aboveground carbon (AGC) from 1988 to 2021.
All datasets are harmonized to 0.25\textdegree{} spatial resolution (ca. 25\,km at the Equator) with annual temporal resolution (see Supplementary Table 1 and Supplementary Fig. 1 for detailed data information and processing steps).
AGC is normally derived from AGB using empirical coefficients (e.g., 0.5)~\cite{brandtSatellitePassiveMicrowaves2018}; therefore, the following content will focus on the estimation of AGB.

To establish the relationship between AGB and multiple vegetation and environmental variables, we classify all predictors as either dynamic or static based on their temporal extent and variability.
Dynamic predictors are time-dependent and hence capture interannual fluctuations when extrapolating the derived spatial relationship to other years. 
Static predictors represent environmental properties that are not time-dependent in our analysis, either because they are also constant in the real world, or because records are too short (such as L-band VOD).

To reconstruct continuous multi-decadal AGB dynamics in the absence of historical 3D structural data, we rely on a carefully curated set of long-term optical and microwave proxies. 
As dynamic predictors, we select the recently released VODCAv2 CXKu-band VOD~\cite{zottaVODCAV2Multisensor2024}, the PKU GIMMS NDVI~\cite{liSpatiotemporallyConsistentGlobal2023}, and GIMMS LAI 4g~\cite{caoSpatiotemporallyConsistentGlobal2023}, alongside tree Plant Functional Types (PFTs) and forest cover fractions derived from ESA CCI Land Cover data~\cite{cciLandCoverCCI2017}.
Crucially, these variables capture complementary biophysical properties.
While optical indices like NDVI reflect canopy density and photosynthetic capacity~\cite{gamon1995relationships}, which are fundamental drivers of carbon accumulation, LAI additionally exhibits a power-law scaling relationship with structural parameters such as tree height~\cite{west1999general,zhang2014estimation}, serving as a useful proxy for vertical forest growth.
Integrating these optical proxies with passive microwave observations (CXKu-band VOD), which are highly sensitive to woody volume and water content~\cite{liuRecentReversalLoss2015}, allows our model to leverage complementary biophysical sensitivities. 
This combination provides a more comprehensive and historically consistent characterization of AGB changes~\cite{liuGlobalLongtermPassive2011}, while the incorporated land-cover metrics supply essential structural constraints on spatiotemporal forest variations.

Since CXKu-band VOD, NDVI, and LAI can have missing observations over frozen or snow-covered surfaces in northern latitudes during winter, we only use growing-season values to ensure representative model inputs.
To maintain consistency across CXKu-band VOD, NDVI, and LAI, we adopt fixed-month windows for latitudinal regions following Zhao et al.~\cite{zhaoChangesGlobalVegetation2018}, instead of defining growing seasons per grid cell.
We calculate the annual mean values of CXKu-band VOD, NDVI, and LAI during the growing season, along with the 95\% percentile of CXKu-band VOD, as it exhibits greater sensitivity to interannual carbon dynamics~\cite{douReliabilityUsingVegetation2023}.
This combination also helps balance the contribution of passive microwave and optical data in modeling.

We employ the user tool of ESA CCI Land Cover to generate PFTs of trees at 0.25\textdegree{} resolution~\cite{harper29yearTimeSeries2023}, and then derive forest cover fractions by summing relevant forest-related land-cover fractions~\cite{besnardGlobalSensitivitiesForest2021}. 
Because the derived LAI data (1988--2020) do not extend to 2021, we use an autoregressive integrated model to project them by one additional year.
Specifically, we implement an ARIMA(1,1,0) model for each grid cell using annual time series data.
For each grid cell, if the annual LAI series contains no missing values, we fit the ARIMA model to the 1988--2020 data and forecast a single value for 2021.
This process is applied independently to each grid cell.
In cases where the time series contains missing data or the model fails to converge, the 2021 value is set to no data.
Similarly, to align PFT and forest cover fraction data (originally from 1992--2021) with our study period of 1988--2021, we apply the same model-based extension for the earlier years.
This short temporal extrapolation is used solely to ensure temporal alignment across model inputs and is not intended for independent interpretation.
Moreover, since other input variables (e.g., VOD and NDVI) have sufficiently long time series, they provide complementary information that helps constrain the model during training, thereby mitigating potential uncertainties introduced by this extrapolation.

As static predictors, we employ the VODCAv2 L-band VOD, the MODIS PAR product (MCD18C2)~\cite{wangNewSetMODIS2020}, the ETOPO 2022 Global DEM~\cite{macferrinEarthTopography20222024}, and geographically encoded coordinates.
L-band VOD offers stronger penetration capability than higher-frequency microwave products (e.g., CXKu-band VOD) and exhibits a stronger relationship with AGB~\cite{wigneronGlobalCarbonBalance2024}.
However, it is only available from 2010 to 2021, hence we treat it as a static predictor for capturing aspects of vegetation density.
Likewise, we use PAR data, available from 2000 to the present, to represent radiation availability.
We aggregate the available L-band VOD (growing-season) data to compute total means and 95\% quantiles, while we use total means for PAR.
The ETOPO 2022 DEM is included as a static predictor due to its near-constant nature; despite being a 2022 product, large-scale topography generally remains unchanged over decadal timescales.
To account for the cyclical nature of the Earth, longitude is transformed into two continuous variables using sine and cosine functions: $\sin(\pi \cdot \text{lon} / 180)$ and $\cos(\pi \cdot \text{lon} / 180)$.

\subsection*{AGC references}\label{reference}

All AGB reference data used in this study are publicly available and either originally provided at, or aggregated to, a spatial resolution of 0.25\textdegree{}, matching the resolution of predictors.
These AGB reference maps are also converted to AGC by applying a fixed conversion factor of 0.5, assuming that approximately 50\% of AGB is composed of carbon~\cite{brandtSatellitePassiveMicrowaves2018}.
Supplementary Table 2 details the data sources.
We employ the ESA CCI AGB maps (version 5.01) as our primary target in model development because they integrate synthetic-aperture radar (SAR) backscatter and spaceborne light detection and ranging (LiDAR) metrics with globally consistent algorithms, yielding greater reliability in woody biomass estimation than many model-based AGB products~\cite{santoroDesignPerformanceClimate2024}.
ESA CCI provides eight AGB maps (one for 2011, plus annual maps from 2015 to 2021) accompanied by standard deviation layers.
However, these maps vary in quality due to observational differences between years. 
The 2017--2021 annual maps are considered equally reliable and supersede any earlier releases~\cite{santoroESACCIBiomass2024}.
AGB changes from year to year are diagnosed by comparing the independent AGB maps from adjacent years.
According to ESA CCI’s quality flags, over 95\% of forested regions exhibit changes labeled as improbable or unrealistic (see Supplementary Fig. 2).
Consequently, to prevent our modeling framework from learning and propagating these spurious temporal artifacts, we neither pool multi-year data into a single model nor explicitly model AGB temporal dynamics.
Instead, we train independent models for each year to reconstruct static AGB patterns (see the following sections for modeling details).

Additional AGB references include global annual maps for 1993--2012 by Liu et al.~\cite{liuRecentReversalLoss2015}, a global forest static map (ca. 2000) by Avitabile et al.~\cite{avitabileIntegratedPantropicalBiomass2016}, pan-tropical annual maps for 2010--2017 by Fan et al.~\cite{fanSatelliteobservedPantropicalCarbon2019}, global annual maps for 2000--2019 by Xu et al.~\cite{xuChangesGlobalTerrestrial2021}, global annual maps for 2011--2023 by Boitard et al.~\cite{boitardAbovegroundBiomassDataset2025}, and a global static map (ca. 2020) by GEDI L4B~\cite{dubayahGEDIL4BGridded2023}.
The products from refs.~\cite{liuRecentReversalLoss2015,fanSatelliteobservedPantropicalCarbon2019,boitardAbovegroundBiomassDataset2025} use VOD-based empirical methods (with Ku\&X-band VOD in Liu et al.~\cite{liuRecentReversalLoss2015}, and L-band VOD in the latter two), whereas the dataset from Avitabile et al.~\cite{avitabileIntegratedPantropicalBiomass2016} merges three earlier maps~\cite{saatchiBenchmarkMapForest2011,bacciniEstimatedCarbonDioxide2012,santoroForestGrowingStock2015}.
The total living biomass maps by Xu et al.~\cite{xuChangesGlobalTerrestrial2021} are derived from microwave and optical remote-sensing imagery with a random forest model; we convert them to AGB by applying the grid-cell-wise AGB-to-belowground biomass ratio from Spawn et al.~\cite{spawnHarmonizedGlobalMaps2020}.
Additionally, GEDI L4B provides a static, relatively sparse global AGB sampling around 2020 (covering missions from week 19 to week 138)~\cite{dubayahGEDIL4BGridded2023}.
We use AGC maps converted from these references, using empirical coefficients (e.g., 0.5) to assess the spatial and temporal reliability of our estimates.
Furthermore, the temporally resolved datasets from refs.~\cite{liuRecentReversalLoss2015, fanSatelliteobservedPantropicalCarbon2019, xuChangesGlobalTerrestrial2021, boitardAbovegroundBiomassDataset2025} allow us to validate our findings on AGC changes across scales discussed in the main text, given their independent time-series coverage and prior validation in other studies.

\subsection*{Regional masks}\label{masks}

\hspace*{1.5em}\textbf{Global forest.} To ensure that AGC change analyses are confined to consistently defined long-term forested regions with a spatial resolution of 0.25\textdegree{} over the past three decades, we follow the forest-masking method described by Besnard et al.~\cite{besnardGlobalSensitivitiesForest2021}.
We derive an annual forest cover fraction from the ESA CCI Land Cover data and calculate its 5th percentile across all periods.
We then retain only those grid cells where this 5th percentile value exceeds a 20\% forest cover threshold~\cite{besnardGlobalSensitivitiesForest2021}.
The resulting forest mask is shown in Supplementary Fig. 16a.
This approach ensures consistency with other ESA CCI-based predictors, such as PFTs and forest cover fractions, as well as ESA CCI AGB. 

\textbf{Forest biomes.} The biome masks are consistent with Xu et al.~\cite{xuChangesGlobalTerrestrial2021}, which classifies global vegetation biomes (i.e., moist tropical, dry tropical \& subtropical, temperate, and boreal biomes) using the MODIS IGBP (International Geosphere-Biosphere Programme) land cover product from 2001.
To ensure consistency with our global forest area, we aggregate the biome mask to a 0.25\textdegree{} resolution and take its spatial intersection with the global forest mask, thereby identifying four distinct forest biomes (Supplementary Fig. 16b).
For forest grid cells that do not overlap with any biome, the nearest biome class is assigned.

\textbf{Deforested regions and intact forests in the Brazilian Amazon.} 
The deforested regions are derived from the INPE PRODES~\cite{institutonacionaldepesquisasespaciaisinpePRODESCoordenacaoGeralObservacao2025}, which conducts satellite-based monitoring of clear-cut deforestation at approximately 30-meter resolution in the Legal Amazon since 1988, producing annual deforestation rates for the region.
PRODES aggregates all deforestation maps from 1988 to 2007 into a single layer, while maps from 2008 onward are provided annually.
To create a long-term deforestation map, we combine deforestation data from 1988 to 2021.
The 30-meter resolution maps are aggregated to a 0.25\textdegree{} grid, and grid cells with a deforestation ratio of 10\% or higher are classified as deforested (Fig.~\ref{fig:fig6}d,e).
To delineate intact forests, we rely on the Intact Forest Landscape (IFL) project. 
An IFL is formally defined as a seamless mosaic of forest and naturally treeless ecosystems that exhibits no remotely detected signs of human activity and covers a minimum area of 500~km$^2$~\cite{potapov2017last}. 
We extract the intact forest boundaries for the Brazilian Amazon from the 2020 IFL map and convert this layer to match our 0.25\textdegree{} grid (Fig.~\ref{fig:fig6}c).

\textbf{Other masks.} Other masks include the regions of countries with the largest forest areas, as reported by the Food and Agriculture Organization of the United Nations (FAO), which are obtained from the Database of Global Administrative Areas (GADM; \url{https://gadm.org/data.html}).
The pan-tropical region is defined as the area within approximately 23.5\textdegree{} north and south of the Equator.
The boundaries of the Amazon Basin and Congo Basin are sourced from ArcGIS Online, (\url{https://worldmap.maps.arcgis.com/home/item.html?id=f2c5f8762d1847fdbcc321716fb79e5a} for the Amazon Basin and \url{https://www.arcgis.com/home/item.html?id=23ed36037ac14b479dd787d7fd418b9e} for the Congo Basin), while the boundary of Indonesia is obtained from GADM.
Rainforest areas within these regions are extracted by intersecting them with the moist tropical forest mask.
All masks are harmonized to a 0.25\textdegree{} grid (Supplementary Fig.16c-e).

\subsection*{Spatially explicit AGC reconstruction}\label{estimation}

We adopt a space-for-time substitution approach (see Supplementary Fig.1), given the scarcity of fine temporal resolution AGC data~\cite{brandtSatellitePassiveMicrowaves2018}.
Rather than relying on empirical VOD--AGB relationships~\cite{liuRecentReversalLoss2015,brandtSatellitePassiveMicrowaves2018,fanSatelliteobservedPantropicalCarbon2019,qinCarbonLossForest2021,fengDoublingAnnualForest2022,yangGlobalIncreaseBiomass2023} or classic machine-learning approaches (e.g., random forest) that capture nonlinear relations on a per grid cell basis~\cite{xuChangesGlobalTerrestrial2021}, we develop a CNN-based, probabilistic deep learning model.
This CNN simultaneously learns the relationship between predictors and AGB targets, as well as the local spatial dependencies (i.e., texture features).
Our network architecture follows Lang et al.~\cite{langHighresolutionCanopyHeight2023}, which has been successfully applied in global canopy-height regression, featuring residual blocks with separable convolutions to stabilize training and enhance computational efficiency.
We employ learnable 3$\times$3 convolutional filters without pooling to retain critical information from limited input datasets. 

To process the global datasets and generate batched training samples for the CNN, we tile the continuous spatial maps into smaller, localized 2D subsets, referred to as patches.
Specifically, we extract 15$\times$15-grid cell patches from all predictor variables (multiple channels) and their corresponding ESA CCI AGB maps.
At a spatial resolution of 0.25\textdegree{}, each patch spans 3.75\textdegree{}$\times$3.75\textdegree{}, enabling the CNN to capture regional interconnections across space.
Because some grid cells in these patches may lack data (e.g., from non-forest areas or missing observations), we exclude these grid cells from the loss function computation.
Given inaccuracies in ESA CCI’s annual AGB changes (Supplementary Fig. 2), we train separate CNNs for each year from 2015 to 2020 (using that year's predictors and ESA CCI AGB), then average their outputs to form an ensemble prediction.
We reserve the 2010 and 2021 data for testing. 
Further hyperparameter details are provided in Supplementary Table 6.
Finally, we apply the ensemble models to predictor patches spanning 1988--2021 to generate spatially explicit AGB estimates, which we convert to AGC via a fixed 0.5 factor.

It is also important to note that the observational data used in our modeling contain spatially and temporally heterogeneous gaps (missing data), primarily due to limitations of optical and passive microwave remote sensing, such as cloud cover, data availability, and sensor-specific constraints.
Although aggregating observations over the growing season helps mitigate much of the missing-data issue, some gaps still persist, and the AGC estimates presented exhibit similar patterns of missing values.
Such data gaps are common in existing AGC datasets derived from remote sensing~\cite{fanSatelliteobservedPantropicalCarbon2019,yangGlobalIncreaseBiomass2023,boitardAbovegroundBiomassDataset2025}.
To ensure the reliability of temporal analyses, we strictly exclude grid cells that have missing values at any point in the time series.
This masking approach ensures that AGC change and flux estimates are based on consistent spatial coverage without the influence of artificial data gaps.
However, because these masked grid cells are omitted from spatial aggregations, this approach inevitably leads to a systematic underestimation of the absolute total AGC stocks in the corresponding regions. 
Nevertheless, since the primary focus of this study is on AGC changes, trends, and fluxes rather than absolute stock inventories, such conservative masking does not affect our main findings.

\subsection*{Predictive uncertainty quantification}\label{uncertainty}

Predictive uncertainty in deep learning can be categorized into aleatoric and epistemic uncertainties~\cite{kendallWhatUncertaintiesWe2017}.
In our context, aleatoric uncertainty stems from noise in both predictor variables and the AGB target, whereas epistemic uncertainty reflects the CNN’s limitations in modeling, and the limited information content of the input data.
We employ deep ensembles with a weighted Gaussian negative log-likelihood (NLL) loss (Equation~\ref{eq:nll-loss}) to jointly capture these two sources of uncertainty~\cite{langHighresolutionCanopyHeight2023}.
Specifically, at each grid cell $i$, we represent the output of model $\theta$ as a conditional Gaussian distribution over AGB target $y_i$, given the input data $x_i$.
We estimate both the mean $\hat{\mu}_{\theta}(x_i)$ and variance $\hat{\sigma}^2_{\theta}(x_i)$ of this distribution~\cite{kendallWhatUncertaintiesWe2017,lakshminarayananSimpleScalablePredictive2017}.
The weight $w_i$ corresponds to the ratio between the ESA CCI AGB value and its associated standard deviation at grid cell $i$, thereby emphasizing grid cells with lower relative uncertainties (i.e., higher reliability in ESA CCI AGB).
To promote stable convergence, we begin training with a weighted mean-square-error loss and then fine-tune with the weighted Gaussian NLL loss:

\begin{equation}
\label{eq:nll-loss}
\mathcal{L}_{\text{NLL}} = \frac{1}{N} \sum_{i=1}^{N} w_i \left( \frac{(\hat{\mu}_{\theta}(x_i) - y_i)^2}{2\,\hat{\sigma}_{\theta}^2(x_i)} + \frac{1}{2}\,\log \hat{\sigma}_{\theta}^2(x_i) \right).
\end{equation}

We train an ensemble of five CNNs for each year from 2015 to 2020, where each network is initialized with a different random seed.
This results in a total of $M=30$ CNN realizations, each producing a predictive mean and variance for AGB.
Following Lakshminarayanan et al.~\cite{lakshminarayananSimpleScalablePredictive2017}, we aggregate these outputs as an equally weighted mixture of Gaussian predictive distributions.
Specifically, for each grid cell $x_i$, we compute the ensemble predictive mean as the average of the component means, and derive the total predictive variance using the law of total variance.
The latter consists of two terms: the average predicted variance across ensemble members, which captures aleatoric uncertainty, and the variance of the predicted means across ensemble members, which reflects epistemic uncertainty.
This yields the final combined mean $\hat{\mu}_*$ and total variance $\hat{\sigma}_*^2$ for each grid cell $i$ (Equations~\ref{eq:mix-mu} and ~\ref{eq:mix-sigma}).

\begin{equation}
\label{eq:mix-mu}
\hat{\mu}_*(x_i) = \frac{1}{M} \sum_{m=1}^{M} \hat{\mu}_{\theta_m}(x_i),
\end{equation}

\begin{equation}
\label{eq:mix-sigma}
\hat{\sigma}_*^2(x_i) = \frac{1}{M} \sum_{m=1}^{M} \left(\hat{\sigma}_{\theta_m}^2(x_i) + \hat{\mu}_{\theta_m}^2(x_i) \right)-\hat{\mu}_*^2(x_i).
\end{equation}

For the temporal analysis of annual AGC stock trends and changes, we quantified uncertainties via a Monte Carlo approach.
For each grid cell and each year, we sampled 1,000 times from the final grid-cell specific Gaussian distribution ($\hat{\mu}_*$ and $\hat{\sigma}_*^2$) to generate an ensemble of AGC estimates.
Summing over all grid cells for each sampled realization yields a distribution of total (or regional) AGC from which we derive the mean and the 95\% uncertainty interval (the 2.5\% quantile as the lower bound and the 97.5\% quantile as the upper bound).

We determine whether above-ground vegetation in a specific region acts as a sink or source based on the sign of the AGC flux estimate.
We further classify a sink or source as insignificant if: (1) the 95\% uncertainty interval of the flux includes zero, or (2) the sign of the AGC flux contradicts the long-term AGC trend.

\subsection*{Calculation of AGC changes, trends, and fluxes}\label{agc_calculation}

To quantify AGC dynamics, we computed a set of complementary metrics at grid-cell, regional, biome, and global scales. Definitions and calculation methods are as follows:

\begin{enumerate}
    \item \textbf{AGC density.} Each grid cell in our estimates represents AGC density, expressed in $\mathrm{MgC\,ha^{-1}}$.
    
    \item \textbf{AGC stocks.} To obtain AGC stocks at larger spatial scales (e.g., region, biome, global), we aggregated grid-cell-level AGC density by multiplying each grid cell’s AGC density by its corresponding area (with latitudinal weighting) and summing up the total AGC across all grid cells within the target region, expressed in PgC ($1~\mathrm{PgC} = 10^3~\mathrm{TgC} = 10^6~\mathrm{GgC} = 10^9~\mathrm{MgC}$). 
    
    \item \textbf{AGC net change.} We define the net change over a specific period as the difference between the AGB values in the last minus the first year of the time window. To ensure comparability across different time spans (e.g., decadal analyses), we divide this total difference by the number of years in the period, reporting it as a mean annual rate of change. We calculate spatially explicit net change maps from AGC density maps and express them in $\mathrm{MgC\,ha^{-1}\,yr^{-1}}$, while regional and global net changes were obtained by aggregating spatial maps, with results reported in $\mathrm{TgC\,yr^{-1}}$. In a given region and time period, a positive AGC net change is interpreted as a carbon sink, while a negative net change indicates a carbon source.
    
    \item \textbf{AGC gross loss.} Where a grid cell shows a decrease in AGC over a specific period, we count the magnitude of the loss as a gross loss. Gross losses at the regional or global scale are the sum of all negative components of the net change map, and are expressed in $\mathrm{TgC\,yr^{-1}}$. A spatially explicit gross loss map was also derived, representing the average loss per grid cell over the period, with units of $\mathrm{MgC\,ha^{-1}\,yr^{-1}}$. Gross gain was calculated using the same approach but retaining only the positive components. In this study, we primarily focus on gross loss, as it is more directly related to the impacts of natural and anthropogenic disturbances.
    
    \item \textbf{AGC trend.} We estimated long-term trends in AGC density or stocks using the Theil–Sen slope estimator and assessed their statistical significance with the modified Mann–Kendall test to account for serial autocorrelation. Unlike net changes, which only consider the first and last years, the trend reflects continuous temporal change across the full period. Grid-cell-level trend maps are reported in $\mathrm{MgC\,ha^{-1}\,yr^{-1}}$, while aggregated trends are expressed in $\mathrm{TgC\,yr^{-1}}$.
    
    \item \textbf{AGC flux.} Calculated as the annual net change in AGC stocks (i.e., $\Delta\mathrm{AGC} = \mathrm{AGC}_{t} - \mathrm{AGC}_{t-1}$) at regional or global scales. Positive values indicate an annual gain in AGC stock, while negative values indicate a loss. Consequently, because AGC flux is derived from the first-difference of consecutive years, the resulting interannual flux time series naturally spans the 33-year period from 1989 to 2021, derived from the foundational 34-year (1988--2021) stock dataset. This metric (reported in $\mathrm{TgC\,yr^{-1}}$ or $\mathrm{PgC\,yr^{-1}}$) is essential for capturing high-frequency variability in the net carbon balance and assessing interannual responses to transient climatic anomalies.
\end{enumerate}

To mitigate satellite measurement uncertainties caused by volcanic aerosol interference and thereby ensure the robustness of our findings, we exclude the years 1991--1992 from analyses of AGC stock trends, and 1991--1993 from analyses of interannual AGC flux variability, following Liu et al.~\cite{liuIncreasinglyNegativeTropical2023}.

\clearpage
\section*{Declarations}
\backmatter

\bmhead{Author contribution}

Z.Q, S.B. and N.B. conceived and designed the study. Z.Q. performed the data collection and computations and analyzed the results. All authors discussed the results. Z.Q. wrote the paper with contributions from all authors.

\bmhead{Competing interests}

The authors declare no competing interests.

\bmhead{Supplementary information}

Supplementary information is available for this paper.

\bmhead{Data availability}

All predictors and AGC reference data are publicly available from earlier studies, the sources of which are described in Methods and Supplementary Information.
The sources of all masks used in this study are provided in Methods.
The Standardized Precipitation Evapotranspiration Index (SPEI) data are publicly available at \url{https://spei.csic.es/spei_database/#map_name=spei03}.
The annual mean global carbon dioxide growth rates are publicly available at \url{https://gml.noaa.gov/ccgg/trends/gl_gr.html}.
The long-term data on deforestation areas in the Brazilian Amazon are publicly available at \url{http://www.obt.inpe.br/OBT/assuntos/programas/amazonia/prodes}.
The temperature and precipitation data are from the Climatic Research Unit (CRU), publicly available at \url{https://crudata.uea.ac.uk/cru/data/hrg/cru_ts_4.09}.
The Multivariate ENSO Index Version 2 (MEI.v2) data are publicly available at \url{https://psl.noaa.gov/enso/mei}.
The AGC estimates produced in this study will be publicly available upon acceptance of this manuscript for publication.

\bmhead{Code availability}

The code for modeling and analysis will be publicly available upon acceptance of this manuscript for publication.

\bmhead{Acknowledgements}

We thank Laibao Liu, Nuno Carvalhais, and Sung-Ching Lee for their suggestions and comments. Z.Q. acknowledges funding from the program of the China Scholarships Council (no.202306860010). N.B. and S.B. acknowledge funding from the Volkswagen Foundation. T.L. acknowledges funding from the National Key R\&D Program of China no.2023YFE0109000. This is ClimTip contribution \#72; the ClimTip project has received funding from the European Union’s Horizon Europe research and innovation programme under grant agreement no. 101137601. This study received support from the European Space Agency Climate Change Initiative (ESA-CCI) Tipping Elements SIRENE project (contract no. 4000146954/24/I-LR).


\clearpage
\bibliography{zotero-references}

\end{document}